\documentstyle[12pt,aaspp4]{article}

\slugcomment{To appear in Astrophys. J. Letters}

\lefthead{EIGENMODE ANALYSIS}
\righthead{Matsubara, Szalay, \& Landy}

\begin{document}
\newcommand{\gsim}{\mbox{\raisebox{-1.0ex}{$\stackrel{\textstyle >}
{\textstyle \sim}$ }}}
\newcommand{\lsim}{\mbox{\raisebox{-1.0ex}{$\stackrel{\textstyle <}
{\textstyle \sim}$ }}}
\newcommand{\gtsima}{$\; \buildrel > \over \sim \;$}
\newcommand{\ltsima}{$\; \buildrel < \over \sim \;$}
\newcommand{\simgt}{\lower.5ex\hbox{\gtsima}}
\newcommand{\simlt}{\lower.5ex\hbox{\ltsima}}
\newcommand{\himpc}{{\hbox {$h^{-1}$}{\rm Mpc}} }
\newcommand{\bfm}[1]{{\mbox{\boldmath $#1$}}}
\newcommand{\bfd}{{\mbox{\boldmath $k$}}}
\newcommand{\bfk}{{\mbox{\boldmath $k$}}}
\newcommand{\bfq}{{\mbox{\boldmath $q$}}}
\newcommand{\bfr}{{\mbox{\boldmath $r$}}}
\newcommand{\bfx}{{\mbox{\boldmath $x$}}}
\newcommand{\bfy}{{\mbox{\boldmath $y$}}}
\newcommand{\bfz}{{\mbox{\boldmath $z$}}}
\newcommand{\bfv}{{\mbox{\boldmath $v$}}}
\newcommand{\sbfk}{{\mbox{\scriptsize\boldmath $k$}}}
\newcommand{\sbfx}{{\mbox{\scriptsize\boldmath $x$}}}
\newcommand{\bfpsi}{{\mbox{\boldmath $\psi$}}}
\newcommand{\bfPsi}{{\mbox{\boldmath $\Psi$}}}
\newcommand{\mPsi}{{\mit\Psi}}
\newcommand{\tth}{\widetilde{\theta}}
\newcommand{\thh}{\theta_{\rm h}}
\newcommand{\tdel}{\widetilde{\delta}}
\newcommand{\Mpc}{{h^{-1} {\rm Mpc}}}
\newcommand{\etal}{et al.~}

\def\pp{\par\parshape 2 0truecm 15.5truecm 1truecm 14.5truecm\noindent}
\renewcommand{\theequation}{\mbox{\rm
{\arabic{section}.\arabic{equation}}}}

\title{COSMOLOGICAL PARAMETERS FROM THE EIGENMODE ANALYSIS OF THE
LAS CAMPANAS REDSHIFT SURVEY}

\author{Takahiko Matsubara\altaffilmark{1,2}, Alexander S. Szalay,}
\affil{Department of Physics and Astronomy,
     The Johns Hopkins University,
     3400 N.Charles Street, Baltimore, MD 21218}
\altaffiltext{1}{Department of Physics, The
        University of Tokyo, Hongo 7-3-1, Tokyo 113-0033, Japan}
\altaffiltext{2}{Research Center for the Early Universe,
     Faculty of Science, The University of Tokyo, Tokyo 113-0033, Japan}

\and
\author{Stephen D. Landy}
\affil{Department of Physics, The College of William and Mary,
Williamsurg, VA 23187-8795}

\begin{abstract}
We present the first results of the Karhunen-Lo\`eve (KL) eigenmodes
applied to real data of the Las Campanas Redshift Survey (LCRS) to
simultaneously measure the values of the redshift-distortion
parameter, $\beta = \Omega_0^{\,0.6}/b$, the linearly extrapolated
normalization, $\sigma_8^{\rm L}$, and the CDM shape parameter,
$\Gamma = \Omega_0 h$. The results of our numerical likelihood
analysis indicate a low value of $\beta = 0.30 \pm 0.39$, a shape
parameter $\Gamma = 0.16 \pm 0.10$, and a linearly extrapolated
normalization $\sigma_8^{\rm L} = 0.79 \pm 0.08$, which are consistent
with a low density universe, $\Omega_0 \simlt 0.5$.
\end{abstract}


\setcounter{equation}{0}

\section{Introduction}

The accurate measurement of cosmological parameters has been a
long-standing challenge for cosmologists. Fortunately, the rapidly
increasing size of redshift surveys is moving the estimation of
many of these parameters out of the shot-noise limited regime.
With these larger data sets, more precise measurements now depend
on correspondingly more sophisticated methods of analysis. For
example, in estimating the power spectrum of galaxy clustering,
one of the greatest challenges is in properly accounting for the
effects of a finite survey geometry and the effects of redshift
distortions on the signal.

The observed power spectrum is a convolution of the true power with
the Fourier transform of the spatial window function of the survey, $
P_{\rm obs}(\bfk)=\int P_{\rm true}(\bfk')|W(\bfk-\bfk')|^2d^3k'$. One
can attempt to deconvolve the true power spectrum or compare to
convolved theoretical spectra, but in either case the survey geometry
limits both the resolution and the largest wavelength for which an
accurate measurement can be obtained.

The standard methods for power spectrum estimation (e.g., Park \etal
1994; Feldman \etal 1994; Fisher \etal 1993) work reasonably well
for data in a large, contiguous, three-dimensional volume, with
homogeneous sampling of the galaxy distribution, and a weighting
scheme optimized for the shot-noise dominated errors. Using these
techniques, nearby wide-angle redshift surveys (CfA, SSRS, IRAS 1.2,
QDOT) yield strong constraints on the power spectrum on scales
approaching $100\Mpc$. Tegmark \etal (1998) provides a detailed
comparison of power spectrum estimation methods in cosmology.

Because the uncertainty in the power spectrum depends on the number of
independent modes at a given wavelength, constraints on larger scales
require deeper surveys. Due to the difficulty of obtaining redshifts
for fainter galaxies and limited telescope time, deep redshift surveys
typically have complex geometry, e.g., deep pencil beams or
slices. Unfortunately, the standard methods are not efficient when
applied to data in oddly-shaped and/or disjoint volumes, or when the
sampling density of galaxies varies greatly over these
regions. Moreover, convolution of the true power with the complex
window function causes power in different modes to be highly
coupled. In other words, plane waves do not form an optimal eigenbasis
for expansion of the galaxy density field sampled by such
surveys. Other intrinsic problems arise due to redshift distortions
and the effects non-linear fluctuation growth.

As a consequence, advanced methods for power spectrum estimation are
needed that optimally weight the data in each region of the survey,
taking into account our prior knowledge of the nature of the noise and
clustering in the galaxy distribution. These methods must also
incorporate the effects of redshift-distortions to produce unbiased
and robust measurements. In this paper we describe a technique that
can take all these effects into consideration, and present the first
results applied to real data. The method employs Karhunen-Lo\`eve (KL)
eigenmodes and is based on the technique outlined by Vogeley and
Szalay (1996) (see also Hoffman 1999), merged together with the
analytic results of redshift distortions in wide angle redshift
surveys by Szalay, Matsubara and Landy (1998). This analysis uses the
largest publicly available redshift survey, the Las Campanas Survey,
LCRS, (Shectman \etal 1996).

\section{Construction of the Eigenbasis}

\subsection{A Short Overview of the Karhunen-Loeve Transform}

In a KL analysis, the survey data is represented as galaxy counts in a
finite number of $N$ cells.  In practice, the data vector $\bfm{d}$ is
defined as
\begin{equation}
   d_i = n_i^{\,-1/2} (c_i - n_i),
   \label{eq1}
\end{equation}
where $c_i$ is the observed number count of galaxies in the $i$-th
cell, and $n_i = \langle c_i\rangle$ is the expected number of
galaxies, based on the number of fibers in the LCRS observation. The
factor $n_i^{\,-1/2}$ whitens the shot noise term (see Vogeley \&
Szalay 1996).  This vector is then expanded over the KL basis
functions $\bfm{\Psi}_n$ as
\begin{equation}
   \bfm{d} = \sum_n B_n \bfm{\Psi}_n.
   \label{eq2}
\end{equation}
The cosmological information is contained in the amplitude and
distribution of the coefficients $B_n$.

The KL eigenmodes are uniquely determined by the following
conditions: (a) Orthonormality,
$\bfm{\Psi}_n\cdot\bfm{\Psi}_m=\delta_{nm}$, and (b) Statistical
orthogonality, $\langle B_n B_m\rangle = \langle B_n^{\,2}
\rangle\delta_{nm}$. This is equivalent to the eigenvalue problem
$ \bfm{R} \bfm{\Psi}_n = \lambda_n \bfm{\Psi}_n$ (Vogeley \&
Szalay 1996), where
\begin{equation}
   R_{ij} = \langle d_i d_j \rangle =
   n_i^{\,1/2} n_j^{\,1/2} \xi_{ij} + \delta_{ij} + \eta_{ij}.
\label{eq4}
\end{equation}
is the correlation matrix calculated for this geometry and choice of
pixelization. $\eta_{ij}$ describes the additional noise terms arising
from systematic effects, like extinction, or total number of fibers in
a given area of the sky.

   Although there is a large degree of freedom in the choice of
pixelization, it is advantageous to choose a pixelization with the
lowest resolution appropriate to the question at hand to reduce
computing time, which is proportional to $N^3$. Since this
research focused on cosmological measurements in the linear
regime, the survey volume was divided into cells about $15 \times
15 \times 40 (\himpc)^3$ in size. The cells are elongated in the
direction of the line-of-sight, to reduce the Finger-of-God
effects. The cell boundaries are based on polar coordinates,
closely following the original tiling of the survey. This resulted
in 1440 and 1503 cells for the Northern and Southern sets of
slices in the LCRS, respectively.

\subsection{Computation of the Correlation Matrix in Redshift Space}

Since the data resides in redshift space, the correlation matrix
must be calculated in redshift space to decompose the signal
properly. The difficulty here is in calculating the cell-averaged
correlation function $\xi_{ij}$ in redshift space, which is used
to construct the correlation matrix $\bfm{R}$. Szalay, Matsubara
\& Landy (1998) derived an analytic expression for the two-point
correlation function in redshift space without using the distant
observer approximation. In this expansion, $\xi^{(s)}$ is given by
\begin{eqnarray}
&&
  \xi^{(s)}(\bfm{r}_1,\bfm{r}_2) =
  c_{00} \xi^{(0)}_0 + c_{02} \xi^{(0)}_2 +
  c_{04} \xi^{(0)}_4 + ...\ ;
\nonumber\\
&&\qquad
  \xi_L^{(n)}(r) = \frac{1}{2\pi^2}
  \int dk\,k^2 k^{-n} j_L(kr) P(k),
  \label{eqa16}
\end{eqnarray}
where
\begin{eqnarray}
&&
  c_{00} =
  1 + \frac23 \beta + \frac{1}{5} \beta^2 -
  \frac{8}{15} \beta^2 \cos^2\theta \sin^2 \theta ,
\label{eqa47}\\
&&
  c_{02} =
  - \left( \frac43 \beta + \frac47 \beta^2 \right)
  \cos 2\theta P_2(\mu) -
  \frac23 \left(\beta - \frac17 \beta^2 +
       \frac47 \beta^2 \sin^2\theta \right) \sin^2\theta,
\label{eqa48}\\
&&
  c_{04} =
  \frac{8}{35} \beta^2 P_4(\mu) -
  \frac{4}{21} \beta^2 \sin^2 \theta P_2(\mu) -
  \frac{1}{5} \beta^2
  \left(\frac{4}{21} - \frac{3}{7} \sin^2 \theta \right)
  \sin^2 \theta,
\label{eqa49}
\end{eqnarray}
and $\beta=\Omega_0^{0.6}/b$, the usual parameter used in relating
velocities to the density field, where $b$ is the bias parameter. The
additional coefficients in a complete expansion ($c_{11}$, $c_{13}$,
$c_{20}$, $c_{22}$) are small enough so that they can be ignored in
this analysis. The geometry of any two points, $\bfm{r}_1$ and
$\bfm{r}_1$ is parameterized by $r = |\bfm{r}_1 - \bfm{r}_2|$,
$\cos(2\theta) = \hat{\bfm{r}}_1 \cdot \hat{\bfm{r}}_2$, and $\mu =
\cos\theta (r_1 - r_2)/r$.

The cell-averaged correlation function is calculated by
numerically integrating the above equations. This is done by an
adaptive Monte-Carlo integration for adjacent pairs of cells, and
by a second-order Taylor approximation for more distant pairs.
\begin{eqnarray}
   \xi_{ij} = \frac{1}{v_i v_j} \int_{v_i} \int_{v_j} dv_i dv_j
   \xi^{(s)}(\bfm{r}_1, \bfm{r}_2). \label{corr}
\end{eqnarray}
A CDM-type power spectrum with $\Gamma = 0.2$, $\sigma_8^{\rm L} =
1.0$, and $\beta = 0.5$ is used to construct the initial KL basis.
This initial choice does not bias any subsequent results, since we
adopted an iterative procedure for our likelihood analysis.

After determining the eigensystem of the matrix $\bfm{R}$, the KL
modes are sorted by descending eigenvalue. The eigenvalues closely
represent the signal-to-noise ratio of each mode. In addition, the KL
modes with large eigenvalues correspond to the larger wavelength
fluctuations (Vogeley \& Szalay 1996). For further analysis we only
use the first $M$ modes ($M<N$). This both reduces the necessary
computations and selects modes where linear theory is more
applicable.

How do we select $M$?  For the essentially two-dimensional
geometry of the LCRS survey the 3D window function in $k$-space is
a very elongated cigar, with the major axis perpendicular to the
plane of the survey, while the 2D window function is extremely
compact (Landy \etal 1996). The KL modes fill the available
$k$-space as densely as possible, given the survey geometry. The
long wavelength KL modes in our truncated basis correspond to the
densely packed 2D modes, thus the cutoff wavelength is
proportional to $M^{-1/2}$, where $M$ is the number of modes in
the truncated set. A true 3D survey would yield better results,
since the cutoff wavelength would scale as $M^{-1/3}$.

On the one hand, one would like to select as many modes as possible,
because then the cosmic variance of the measured parameters is
smaller, due to the averaging over a larger set of random numbers. On
the other hand, including lower signal-to-noise modes not only brings
us closer to non-linear scales but also dilutes the signal-to-noise.
It is non-trivial to balance this issue of cosmic variance versus
non-linearity especially given the complex geometry of the LCRS. A
natural method is to inspect the sorted window functions to see the
relevant scales of each mode to single out the inappropriate scales.
However, we found the resulting window functions had much too complex
shapes for that purpose. This is because the Las Campanas Survey has a
complex geometry and a selection function which varies from field to
field and consequently makes eigenmodes form complex window functions
in terms of Fourier modes. In this paper, we choose the maximum number
of modes $M$ which reproduces reasonable estimates of cosmological
parameters in analyzing mock catalogs drawn from $N$-body simulations
in which true values of parameter are known.

\section{Likelihood Analysis}

The likelihood function (LF) is obtained from the expression
\begin{eqnarray}
   {\cal L} \propto |\det\bfm{C}_{\rm model}|^{-1/2}
   \exp\left[
      -\frac12 \bfm{B}^T \bfm{C}_{\rm model}^{\,-1} \bfm{B}\right],
    \label{eq5}
\end{eqnarray}
where $\bfm{C}_{\rm model}$ is the covariance matrix computed from our
theoretical model hypotheses for a set of parameters $\Pi (\beta,
\sigma_8^{\rm L}, \Gamma)$, rotated to the KL basis. This matrix is
very close to diagonal. For $i,j = 1,\ldots,M$,
\begin{equation}
    (C_{\rm model})_{ij} = \langle B_i B_j \rangle_{\rm model}=\bfm{\Psi}_i
    \bfm{R}_{\rm model} \bfm{\Psi}_j,
   \label{eq6}
\end{equation}

In practice, the correlation matrix can be expressed as a linear
combination of several matrices, proportional to powers of $\beta$ and
$\sigma_8^{\rm L}$. In this analysis, $\sigma_8^{\rm L}$ is always
understood as the linear amplitude of the fluctuation spectrum. The
shape of the respective correlation functions only depends on
$\Gamma$, therefore we computed the matrices for each value of
$\Gamma$, but then computed their linear combinations for the various
values of $\beta$ and $\sigma_8^{\rm L}$. The calculations are still
quite computationally intensive, and were only possible by using
efficient numerical algorithms. Details of our numerical analysis will
be described in a longer, more technical paper (Matsubara \etal 1999).
This paper will also discuss other, higher-dimensional
parameterizations of the power spectrum, like the use of band-power
amplitudes.

Our original fiducial choice of parameters determined the initial KL
basis. After the maximum of the LF is determined, the KL basis is
recomputed at that point, and the likelihood analysis repeated. In the
subsequent section, the results are reported as both LF contours, and
marginalized one-dimensional LF.

\section {Analysis of N-body Simulations}

The N-body simulations were kindly supplied by C. Park and are
the same ones that have been used in earlier analyzes of the LCRS
(Landy \etal 1996, Lin \etal 1996). The simulation is an open CDM
model with $h = 0.5$, $\Omega_0 = 0.4$, and $b = 1$. The model was
normalized so that $\sigma_8=1$. Thus, in this analysis, $\beta =
0.577$, $\Gamma = 0.2$. Determination of $\sigma_8^{\rm L}$ from the
data is problematic due the nonlinear effects and the
finiteness of the volume.

\begin{figure}
\epsscale{1.0} \plotone{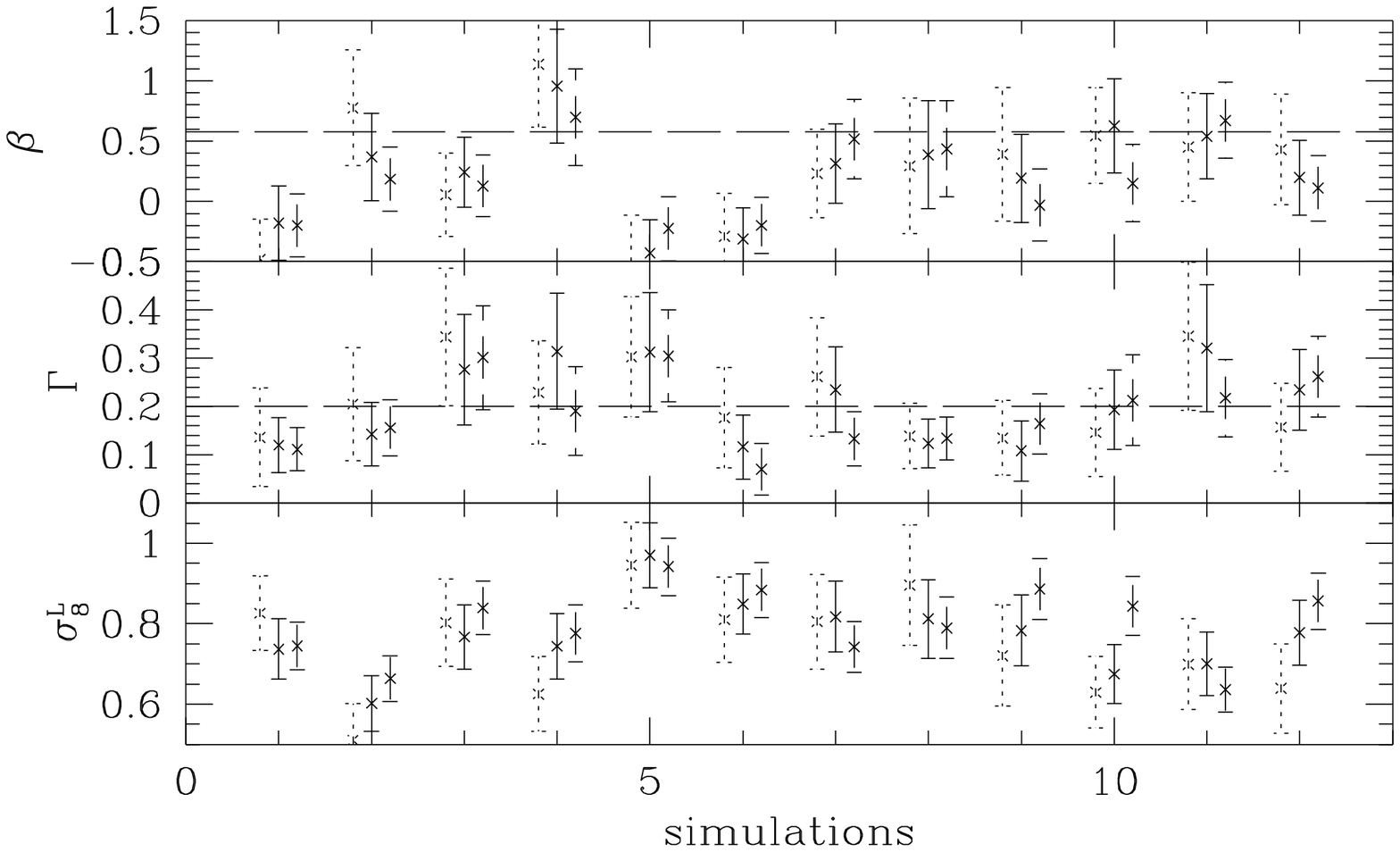} \figcaption[fig1.eps]{The estimation
of parameters in $N$-body simulations. Horizontal axis indicate the
different realization of mock catalog. The true values in the
simulations are shown by long-dashed horizontal lines for $\beta$ and
$\Gamma$. The dotted lines are for $M=100$, solid lines for $M=150$,
and dashed lines for $M=200$.
\label{fig1}}
\end{figure}

The three-dimensional LF for $M = 100, 150, 200$ was computed in a
$21^3$ grid in parameter space $\Pi$. The LF was then marginalized
with respect to each parameter, $\beta, \Gamma, \sigma_8^{\rm L}$. The
resulting discrete LF was fitted by Gaussian curve, in which the
center and the variance are identified with our estimate of the
parameters and its 1$\sigma$ error bars. In Figure \ref{fig1}, the
resulting estimates are plotted.
\placefigure{fig1}
Experimenting showed that iterating
the basis did not change the estimation, thus for the model data we
did not iterate in this figure.

The N-body results show that there is excellent agreement with the
shape parameter $\Gamma$, fairly good agreement with $\beta$. It is
difficult to determine what fiducial value should be used for
$\sigma_8^{\rm L}$ in the mock catalogs for comparison. The major problem
arises from the fact that the small-scale resolution of the analysis,
$\sim20 \himpc$, is over twice the scale of non-linear clustering.
This makes direct analytical calculations problematic since we never
truly sample $\sigma_8$. If it is assumed that the analysis here is
accurate, $\sigma_8^{\rm L}$ is expected to be under-estimated by about
$15\%$ of the true $\sigma_8$.

The number of modes $M$ to use in likelihood analysis mainly
affects the estimation of the error bars. Thus we can decide from
Figure \ref{fig1}, which number $M$ should be used in analyzing the
actual LCRS data. One can see that a choice of $M=150$ is
reasonable for the parameter estimation.

\section {Discussion of Results from the LCRS}

\begin{figure}
\epsscale{1.0} \plotone{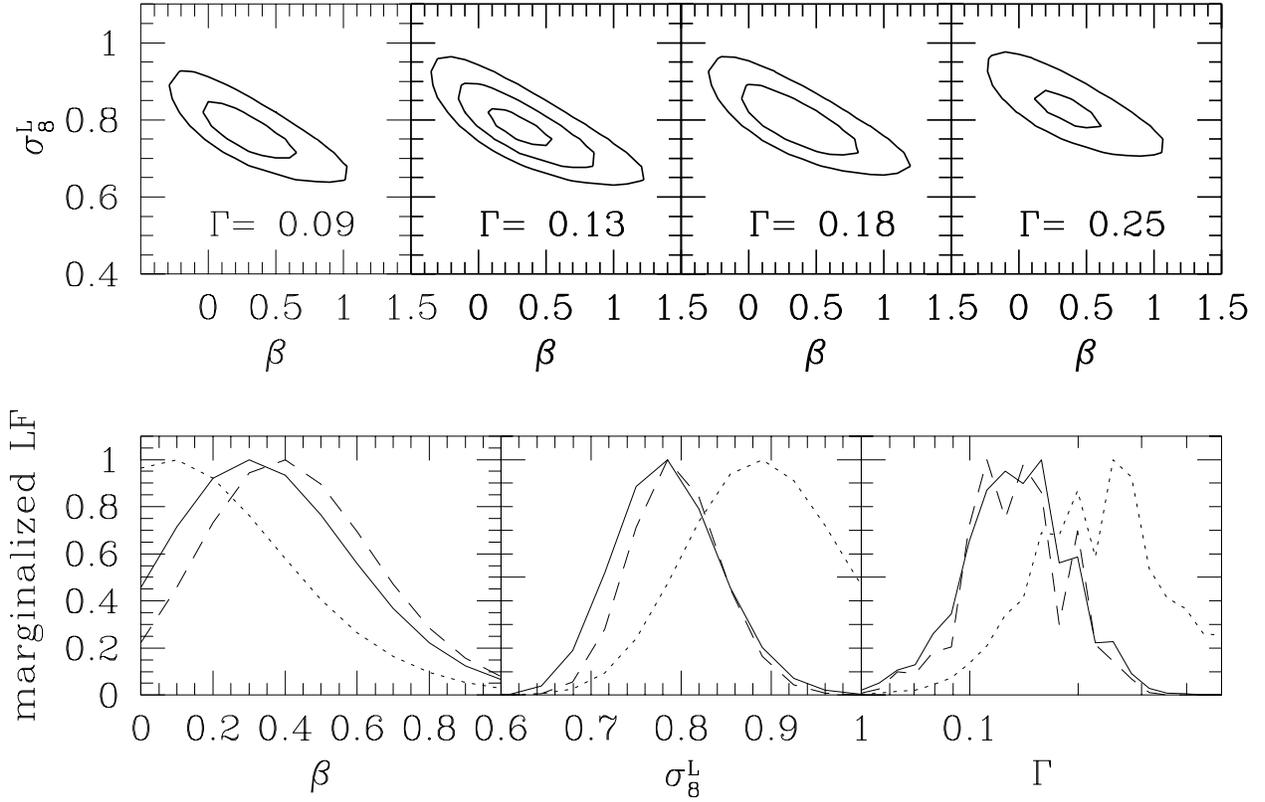} \figcaption[fig2.eps]{Likelihood
function of LCRS data. {\em Upper panel}: Three-dimensional likelihood
function. Four sections of fixed $\Gamma$ are plotted as a contour
map, in which $1\sigma$, $2\sigma$, and $3\sigma$ confidence levels are
shown. {\em Lower panel}: Marginalized likelihood functions of each
parameter. Number of KL modes used in likelihood analysis is varied.
Dotted lines are for $M = 100$, solid lines for $M = 150$, and dashed
lines for $M = 200$.
\label{fig2}}
\end{figure}

We have calculated the LF in the 3D parameter space for both the
Northern and Southern samples separately, then for the combined set.
We used a truncated base with $M = 150$ that was determined from
experience with the N-body simulations as described above. In Figure
\ref{fig2}, several sections of the three-dimensional LF and the
marginalized LF from the actual LCRS data are shown.
\placefigure{fig2}
In Table
\ref{Tab1}, the best fit model parameters are summarized.
\placetable{Tab1}
The number
of iterations is three. The results of the first two iterations are
also consistent with the final estimation in Table \ref{Tab1}.

$\sigma_8^{\rm L}$ is consistent with expectations from simulations
and indicates that true $\sigma_8$ is approximately one. The estimate
of the shape parameter, $\Gamma = 0.16 \pm 0.10$, is somewhat lower
than that found in other analyzes although consistent within errors as
derived from the simulations. For example, Feldman \etal (1994) find
$\Gamma=0.20$ and Landy \etal (1996) find $\Gamma=0.24$ below $75
\himpc$. If $b \approx 1$, the parameter values in the Table
\ref{Tab1} indicate a low value of $\Omega_0\approx \beta^{1.67}
\simlt 0.5$.

Here, it should be noted the limitations of these results with respect
to a CDM three-dimensional parameterization of the shape and amplitude
of the power spectrum. Earlier work by Broadhurst \etal (1989) and
Landy \etal (1996) have shown a perturbation of the power spectrum on
$100 \himpc$ scales. This 'bump' in the power spectrum cannot be
resolved by such a parameterization and would lead to an
under-estimation of $\Gamma$ as the fit finds an average shape. The
other models of the large-scale structure, including PIB or defect
models, can be also studied using the present formalism with
appropriate parameterizations of power spectrum.

The method we have developed here can be straightforwardly applied to
redshift data of any geometry and of any selection function. By
restricting our analysis to the large-wavelength modes, our method
does not depend much on correction for nonlinear effects. The error
bars of the results clearly show the advantages offered by surveys of
larger volume and more isotropic geometry, like SDSS, which would
increase the number of independent large scale modes substantially.

\acknowledgements

We would like to thank the LCRS collaboration for
creating the largest publicly available redshift survey to date.  TM
was supported by JSPS Postdoctoral Fellowships for Research
Abroad. SL would like to recognize support from the
Jeffress Memorial Trust and NSF Grant AST-9900835, AS has been
supported by NSF AST-9802980 and NASA NAG5-3503.

{}

\newpage


\bigskip

\newpage

\begin{table}[h]
\caption{Best fit models from the projected likelihood functions
with $M=150$. Error ranges correspond to 1 $\sigma$ of the
likelihood functions.} \label{Tab1}
\begin{center}
\begin{tabular}{cccc}
  \hline\hline
  Sample & $\beta$ & $\Gamma$ & $\sigma_8^{\rm L}$ \\
  \hline\hline
  North &
  $0.24 \pm 0.42$ & $0.16 \pm 0.11$  & $0.77 \pm 0.09$ \\
  South &
  $0.36 \pm 0.37$ & $0.17 \pm 0.08$  & $0.81 \pm 0.09$ \\
  Combined &
  $0.30 \pm 0.39$ & $0.16 \pm 0.10$  & $0.79 \pm 0.08$ \\
  \hline
\end{tabular}
\end{center}
\end{table}


\begin{thebibliography}{}

\bibitem[Broadhurst \etal (1990)]{bro90} Broadhurst,T.J, Ellis, R.S.,
	Koo, D.C. \& Szalay, A.S. 1990, Nature, 343, 726.
\bibitem[Feldman, Kaiser, \& Peacock (1994)]{fel94} Feldman, H.A.,
    Kaiser, N., \& Peacock, J.A. 1994, \apj, 426, 23.
\bibitem[Fischer \etal (1993)]{fis93} Fisher, K.B, Davis, M., Strauss, M.A.,
    Yahil, A., \& Huchra, J.P. 1993, \apj, 402, 42.
\bibitem[Hoffman (1999)]{hof99} Hoffman, Y. 1999, {\rm in} Cosmic
Flows. Towards an Understanding of Large-Scale Structure, Workshop
Victoria B.C., July 1999, eds. S. Courteau, M. Strauss, and J. Willick
(astro-ph/9909158)
\bibitem[Landy \etal (1996)]{lan96} Landy, S.D., Shectman, S.A., Lin H.,
    Kirschner, R.P., Oemler, A.A. \& Tucker, D. 1996, \apjl, 456, L1.
\bibitem[Lin \etal (1996)]{lin96} Lin, H., Kirshner, R. P., Shectman, S. A., 
    Landy, S. D., Oemler, A., Tucker, D. L., \& Schechter, P. L.
    1996, \apj, 471, 617
\bibitem[Matsubara \etal (1999)]{mat99} Matsubara, T., Szalay, A. S.
\& Landy, S. D. 2000, in progress.
\bibitem[Park \etal (1994)]{par94} Park, C., Vogeley, M. S., Geller, M. J.,
    \& Huchra, J. P.  1994, \apj, 431, 569.
\bibitem[Shectman \etal (1996)]{LCRS96} Shectman, S.A., Landy, S.A.,
    Oemler, A., Tucker, D.,Lin, H., Kirschner, R.L. \& Schechter,
    P.L. 1996, \apjs, 470, 172.
\bibitem[Szalay \etal (1998)]{sza98} Szalay, A. S., Matsubara, T.  \&
    Landy, S. D. 1998, \apj, 498, L1
\bibitem[Tegmark \etal (1998)]{teg98} Tegmark, M., Hamilton, A.J.S.,
    Strauss, M., Vogeley, M.S. \& Szalay, A.S. 1998, \apj, 499, 555.
\bibitem[Vogeley \& Szalay (1996)]{vog96} Vogeley, M. S. \& Szalay, A. S.
    1996, \apj, 465, 34.

\end{thebibliography}
\end{document}